\documentclass[12pt]{article}

\usepackage{graphicx}
\usepackage{amsmath,amssymb}
\usepackage{bm}
\allowdisplaybreaks
\begin{document}
\baselineskip=0.7cm

\newcommand{\e}{{\rm e}}
\newcommand{\Tr}{{\rm Tr}}
\renewcommand{\figurename}{Fig.\@}
\renewcommand{\thesection}{\arabic{section}.}
\renewcommand{\thesubsection}{\arabic{section}.\arabic{subsection}}
\makeatletter
\def\section{\@startsection{section}{1}{\z@}{-3.5ex plus -1ex minus 
 -.2ex}{2.3ex plus .2ex}{\large}} 
\def\subsection{\@startsection{subsection}{2}{\z@}{-3.25ex plus -1ex minus 
 -.2ex}{1.5ex plus .2ex}{\normalsize\it}}
\makeatother
\makeatletter
\def\lesim{\mathrel{\mathpalette\gl@align<}}
\def\gtsim{\mathrel{\mathpalette\gl@align>}}
\def\gl@align#1#2{\lower.7ex\vbox{\baselineskip\z@skip\lineskip.2ex%
  \ialign{$\m@th#1\hfil##\hfil$\crcr#2\crcr\sim\crcr}}}
\makeatother

\def\thefootnote{\fnsymbol{footnote}}

\begin{flushright}
hep-th/0609007\\
UT-KOMABA/06-8\\
September 2006
\end{flushright}
\vspace{0.3cm}
\begin{center}
\Large 
Holography 
of  Wilson-Loop Expectation Values \\
with Local Operator Insertions 

\vspace{0.7cm}

\normalsize
 \vspace{0.4cm}
Akitsugu {\sc Miwa}
\footnote{e-mail address:\ \ {\tt akitsugu@hep1.c.u-tokyo.ac.jp}}
and 
Tamiaki {\sc  Yoneya}
\footnote{
e-mail address:\ \ {\tt tam@hep1.c.u-tokyo.ac.jp}}

\vspace{0.3cm}

{\it Institute of Physics, University of Tokyo\\
Komaba, Meguro-ku, Tokyo 153-8902}

\vspace{1cm}
Abstract
\end{center}
\vspace{0.4cm}
We study the expectation values of Wilson-loop 
operators with the insertions of local operators 
$Z^J$ and $\overline{Z}^J$ for large $J$ from the bulk viewpoint of 
AdS/CFT correspondence. 
Classical solutions of 
strings attached to such 
deformed Wilson loops at the conformal  boundary 
are constructed 
and are applied to the computation of 
Wilson-loop expectation values. 
We argue that in order to have such solutions for general 
insertions at finite positions in the base spacetime of 
the gauge theory, it is crucial to interpret the holographic 
correspondence in the semi-classical picture as a tunneling 
phenomenon, as has been previously established for holographic 
computations of correlators of BMN operators. This also requires 
to use the Euclideanized AdS background and Euclidean super Yang-Mills 
theory.

\newpage
\section{Introduction}
In the applications of the conjectured  
AdS/CFT (or gravity/gauge) correspondence \cite{AdS/CFT,gkpw}, Wilson-loop expectation values have been a focus of much interest.  Besides their role 
as an important probe for studying  
phase structures of gauge theories, the 
Wilson-loop operators have long been regarded as a clue to  
possible string picture for gauge theories. 
With the advent of the AdS/CFT correspondence, 
Wilson loops in general can be associated with the 
 world sheet of strings in the bulk AdS spacetime, 
in such a  way that the boundary of the string world sheets 
coincides with the locus of a given Wilson loop located on 
a holographic screen at the conformal boundary of the 
AdS background. 
Using this picture initiated in \cite{AdSWilson}, 
a lot of interesting results have been 
reported.
Most of such results obtained from 
the viewpoint of bulk string theory should be regarded as 
predictions of the AdS/CFT correspondence, and direct 
checks of corresponding results on the gauge-theory side 
have been difficult because they usually require 
genuine nonperturbative calculations. In certain special cases, 
such as circular (or straight-line) loops 
\cite{Berenstein:1998ij,dgo}
which can be constrained by (super)conformal symmetries,  
some pieces of nontrivial evidence for the agreement of 
both sides have been obtained. For example, it has been shown 
\cite{es,dg,sz}
 that a ladder-graph approximation gives results which are 
consistent with predictions from the string picture in the bulk. 
It is important to extend such correspondence to other cases. 

A first extension to be considered along this line would 
be various small deformations of loops from circle, 
although it is still difficult to perform such calculation 
on the gauge-theory side, since we cannot expect 
that ladder-type approximations continue to be 
valid for general small deformations. 
Another interesting case would be those with various 
insertions of local operators along the Wilson line.\footnote{
Of course we can also understand these local operator insertions 
as deformations of Wilson loop.
See for example \cite{Polyakov}.
} 
In particular, 
we can consider 
insertions of local operators with large R-charge 
angular momentum $J$  
such as $Z(x)^J$ and their cousins, ``spin-chain operators".  
Such operators can 
preserve covariance under an SL(2, R) part of  conformal
transformations.\footnote{
For other cases of Wilson-loop correlators involving 
 large-$J$ local operators, see \cite{zarem}.
}

In a recent paper \cite{dk},
a nice discussion of such Wilson loops  has been given. 
On the bulk side, the authors discussed the spectrum of 
strings in the large $J$ limit. The insertions of the 
local operators are interpreted as 
nonzero  $S^5$-angular momentum density on the 
world sheet 
which is essentially concentrated 
at the center of the AdS$_5$ background in their treatment. 
The special role of the AdS center follows from
 a familiar discussion of the BMN limit \cite{bmn} 
corresponding to $Z^J$, 
while the (doubled) Wilson lines are still located at the 
conformal boundary without any deformation. The absence of  
deformation on the conformal boundary has been 
interpreted as signifying the situation where 
the positions of operator insertions are sent to 
infinity with an infinitely large loop. 

In the context of bulk string picture, however, it is clearly 
more natural to 
consider string world sheets with nontrivial deformations  
corresponding to the insertions of local composite operators at 
generic finite points 
on the conformal boundary. We  would then be able to 
directly evaluate the expectation values for 
arbitrary circular loops using such string 
solutions by following the standard 
bulk interpretation of Wilson-loop expectation values. 
The purpose of the present note is to demonstrate how to 
achieve such a picture in a simplest nontrivial setting 
and to give some remarks relevant 
to this question, in hope of providing a basis for 
further systematic investigations of Wilson-loop 
expectation values with more complicated configurations 
of local operator insertions.  In particular, we will point out that 
a `tunneling' picture which has been developed 
in \cite{dsy,yy} 
in order to reconcile the BMN limit with the 
spacetime holographic picture is crucial for this purpose. 
This approach has led to a natural prescription \cite{dy}
for a holographic correspondence between 
OPE coefficients of the conformal gauge theory to a 
particular 
3-point vertex of string-field theory. 
Establishing such correspondence beyond mere 
comparison of the spectrum has been difficult  in the 
usual approach which does not take into account the 
tunneling picture. 

\section{Tunneling picture and bulk string solutions}
The Wilson-loop operator we consider is 
 \begin{equation}
W[C; \vec{x}_+, \vec{x}_-]=\Tr\Big[{\cal P}
\exp \Big(\oint_C  ds [iA_{\mu}(\vec{x}(s))
\dot{x}^{\mu}
+\sqrt{(\dot{\vec{x}})^2}\phi^i(\vec{x}(s))\theta^i]\Big)
 Z(\vec{x}(s_1))^J\overline{Z}(\vec{x}(s_2))^J
\Big]
\label{dwilson}
\end{equation}
where $\vec{x}_+=\vec{x}(s_1), \, \vec{x}_-
=\vec{x}(s_2)$ 
are the positions of insertions and the contour $C$ 
is assumed to be a circle.  We denote by $\vec{x}=
\{x^{\mu}\}$ 
the 4-dimensional base spacetime coordinates. Other notations 
are standard. For simplicity, we assume that 
the direction of the unit vector $\theta^i$ along the $S^5$ 
is fixed to be the 4-th direction, $\phi^i\theta^i=\phi^4$ and 
that the direction of the angular momentum is 
in the $5$-$6$ plane, $Z=(\phi_5+i\phi_6)/\sqrt{2}$. 
If we wish to study the behavior of the vacuum 
expectation value of this operator from the viewpoint 
of string theory in the bulk, 
the first task is to obtain an appropriate classical 
string configuration such that, as we approach the 
conformal boundary, 
it reduces to the circle $C$ with the 
momentum density along $C$ being concentrated at 
the positions of insertions. We choose the Polyakov-type 
action in the conformal gauge for world-sheet parametrization. 
Also for the target spacetime, we use the Poincar\'{e} 
coordinates for the AdS part with the metric
\begin{equation}
ds^2=R^2\Big[
\frac{dz^2+d\vec{x}^2}{z^2}  +\cos^2\theta (d\psi)^2 +(d\theta)^2 +\sin^2\theta d\tilde{\Omega}_3^2\Big]. 
\end{equation}
For the $S^5$ part, the 4-th direction pointing toward the north pole corresponds to 
$\theta={\pi/2}$ and the angle in the 5-6 plane 
at $\theta=0$ is $\psi$. The angular momentum 
along $\psi$ is then 
\begin{equation}
J=\frac{R^2}{2\pi\alpha'}
\int d\sigma \cos^2\theta(\tau, \sigma)\dot{\psi}(\tau, \sigma) .
\end{equation}

Assume that 
the world-sheet time coordinate $\tau$ is chosen such that 
$\psi(\tau, \sigma)=\tau$.  
The equation of motion then 
requires that $\theta$ is $\tau$-independent. 
The density of the angular momentum with 
respect to $\sigma$ cannot exceed a finite constant 
$R^2/2\pi\alpha'$. 
In order for the R-charge angular momentum to be localized with respect to the target spacetime, the range of $\sigma$ where $\cos\theta\sim  1, 
\vec{x}{\, '}\sim 0, z'\sim 0$  are satisfied must be infinitely large. 
For the $S^5$ part, the solution satisfying the 
criterion is easily found as in \cite{dk}
to be 
\begin{equation}
(\cos \theta, \psi)=(\tanh \sigma, \tau) 
\label{s5part}
\end{equation}
which also satisfies the Virasoro constraint in the 
form
$
 (\dot{\psi})^2 \cos^2 \theta + (\theta')^2=1, \, \, 
 \dot{\psi}\psi' \cos^2 \theta + \dot{\theta}\theta'=0. 
$
The region where the R-charge momentum is localized 
is where $\sigma\rightarrow \infty$. Thus, in the limit of large $\sigma$,  
the AdS coordinates must also be constant with respect to 
$\sigma$ at least when we consider the 
near-boundary region $z\sim 0$. 
Hence the Virasoro-Hamiltonian constraint  
in this region reduces to 
\begin{equation}
\frac{\dot{z}^2 +\dot{\vec{x}}^2}{z^2}+1=0.
\label{minkowconstr}
\end{equation} 
This shows that for the existence of real solution, 
we have to require at least that $\dot{\vec{x}}$ is timelike. 
The equation of motion $\partial_{\tau}\Big({\partial_{\tau}\vec{x}\over 
z^2}\Big)-\partial_{\sigma}\Big({\partial_{\sigma}\vec{x}\over z^2}
\Big)=0$ 
 for $\vec{x}$ reduces, under the same condition, to 
$\dot{\vec{x}}/z^2=$ constant. 
We choose the scale of this integration 
constant by introducing a parameter $\ell$ 
such that  $\dot{\vec{x}}^2/z^4=-\ell^{-2}$. 
However, we then have always $z^2\ge \ell^2$ 
since (\ref{minkowconstr}) is now equivalent to $\ell^2\dot{z}^2=
z^2(z^2-\ell^2)$. 
Since we are seeking for a 
string configuration of which the deformation 
corresponding to local operator insertions 
reaches the conformal boundary 
$z=0$, this is not acceptable. 
Hence, there exists {\it no} desired classical solution which 
reaches the conformal 
boundary even if we assume that the loop $C$ extends in the 
time-like direction. The situation \cite{dsy,yy} 
is the same as in the {\it usual}  
formulation of the BMN limit 
for which it is not possible to directly apply the 
GKPW prescription for the same reason as we encounter here 
that the pp-wave trajectory with 
Lorentzian metric does not reach the conformal boundary. 
Hence, if we remain in the Lorentzian approaches 
such as in \cite{dk}, we cannot achieve the desired picture.

The resolution of this puzzle is quite simple. 
We would like to refer the reader to \cite{dsy,yy}\footnote{
 In the present note, we give only a somewhat sketchy explanation for the
necessity of the tunneling picture to avoid too much repetition.
For an up-to-date review including other references,
we strongly recommend the reader to consult the second ref. in [14]
which also contains a brief discussion on the solution (7)
discussed below.
}
for detailed discussions on this matter. 
In essence, we have to 
take into account that, from the viewpoint of semi-classical 
approximation, the holographic correspondence 
between bulk and conformal boundary is actually 
a tunneling phenomenon as could be inferred 
from the beginning of AdS/CFT correspondence 
embodied in the famous GKPW relation. Namely, we have to 
study the tunneling region $z^2(z^2-\ell^2)\leq 0$. 
As is familiar in elementary quantum mechanics, we perform 
the Wick rotation $\tau
\rightarrow -i\tau$ for the world-sheet time. We 
are then forced to consider Euclideanized 
AdS (EAdS) by assuming that $\vec{x}$ is now a 4-vector with 
Euclidean signature, in order to keep the relative sign 
between $z^4$ and $z^2\ell^2$ against  
the Wick rotation 
of the world-sheet time coordinate. Of course, if 
we started from a loop extending to space-like directions, 
only the Wick rotation with respect to $\tau$ would have been 
sufficient. 
However, when we consider the light-cone 
quantization of strings around 
the classical solution, the Wick rotation with respect to 
the target time direction would be very important irrespectively 
of the directions of the loop.

The Hamiltonian constraint now reduces to $\ell^2\dot{z}^2=
z^2(\ell^2-z^2)$, and the solution of the equation of motion 
 is obtained as 
\begin{equation}
z=\ell /\cosh \tau, \quad x_4=\ell \tanh \tau
\label{tunnel}
\end{equation}
which is nothing but a semi-circle 
$z^2+x_4^2=\ell^2$ in the two-dimensional section $(z, x_4)$ 
and  
reaches the conformal boundary as $\tau
\rightarrow \pm \infty$. 
We have chosen a trajectory lying in the  4-th direction
of the EAdS spacetime. Note that the integration constant 
$\ell$ gives the distance $|\vec x_+ - \vec x_-|=2\ell$ between 
 the two local operator insertions on the boundary.
It should also be noted that as a consequence
of the above Wick rotation we have to rotate the angle
coordinate along the 5-6 plane
simultaneously as $\psi \rightarrow -i\psi$ to keep the value
of R-charge angular momentum $J$ intact.
Thus the 10-dimensional spacetime as a whole is of Lorentzian
signature effectively. This is why we can still perform a
Penrose-type limit in the tunneling picture.
Eq.(6) is the same geodesic trajectory as the
one used for discussing the correlators of the BMN operators and
the associated `holographic' string field theory \cite{dsy, dy}.

 It is now evident that in order to 
have the world-sheet  configuration 
corresponding to our Wilson-loop operator 
with local-operator insertions we have to 
rely on the tunneling picture.  
As in the case \cite{dsy} of the BMN limit, 
the world-sheet time variable $\tau$ 
in (\ref{tunnel}) 
can be identified with the Euclidean time parameter  $\tau$ of 
the global coordinates for EAdS background, 
$
R^2[
\cosh^2\rho (d\tau)^2 +(d\rho)^2 +\sinh^2\rho d\Omega_3^2]
$ 
in terms of which the trajectory is nothing but the 
geodesic at $\rho=0$ which does reach the 
boundary in the limit of large $|\tau|$. 
The full string world-sheet will be described 
as a trajectory of an open string propagating from boundary 
to boundary such that the position of localized R-charge 
momentum traverses along this geodesic of the EAdS background. 
The desired solution can be easily obtained by going to the global EAdS metric 
and then return again to the Poincar\'{e} metric. 
In the case of straight-line Wilson loop,  the solution 
in the global EAdS metric is in fact 
related to the solution discussed in \cite{dk} by our Wick rotation.

Choosing the 
circle $C$ on the boundary in the $x_3$-$x_4$ plane and the 
points of the insertions on the $x_4$ axis, we find
 that the full world sheet is described by two 
patches designated by suffix $\pm$, given as 
\[
(z(\tau, \sigma), x_3(\tau, \sigma), x_4(\tau, \sigma))_{\pm}
\]
\begin{equation}
=
\ell (
\frac{\sinh \sigma}{\cosh \sigma\cosh \tau \pm \alpha}, 
\pm \frac{\sqrt{1-\alpha^2}}{\cosh \sigma \cosh \tau \pm \alpha}, 
\frac{\cosh \sigma \sinh \tau}{\cosh \sigma \cosh \tau \pm \alpha}
). 
\label{sol1}
\end{equation}
In addition to the integration variable $\ell$ appeared already 
to describe the distance of insertion points, we 
introduced another parameter $\alpha$ which gives the radius, 
$r=\ell / \sqrt{1-\alpha^2}$, of the 
circle $C$ for a given $\ell$. 
As a matter of course, solutions with 
circles of different 
sizes on the conformal boundary 
are related by conformal transformations in the bulk EAdS 
background. For our later purpose, it is most convenient 
to keep these two integration constants explicitly. 
The ranges of the world-sheet coordinates are 
$0\le \sigma \le \infty$ and $-\infty \le \tau \le \infty$ for both 
patches. 
The trajectory (\ref{tunnel}), along which
 the R-charge angular momentum is concentrated 
and the two patches are sewn together, 
corresponds to the limit $\sigma\rightarrow \infty$ in 
conformity with the above discussion of localized angular 
momentum. 

On the other hand, the limit 
$\sigma \rightarrow 0$ gives 
\begin{equation}
(z(\tau, 0), x_3(\tau, 0), x_4(\tau, 0))_{\pm}
=\ell (0, \pm \frac{\sqrt{1-\alpha^2}}{\cosh \tau \pm \alpha}, 
\frac{\sinh \tau}{\cosh \tau \pm \alpha})
\end{equation} 
which corresponds to the circle $C$ on the boundary whose center is 
located on the $x_3$ axis.  
The points of local operator insertions are 
$\vec{x}_{\pm}=\ell(0, 0,\pm1) $,  as we 
obtain by taking the limits $\tau\rightarrow 
\pm \infty$, respectively, in either  of these expressions 
($\sigma=0$ or 
$\sigma\rightarrow \infty$). 
For the $S^5$ part, we can adopt the same form as 
(\ref{s5part}) with the understanding of a double 
Wick rotation. 

Figs. 1-3 exhibit the equal-$\tau$ and-$\sigma$ lines on the full world sheet 
with respect to the bulk target space.   
In particular, in the case of the straight-line loop ($\alpha=1$, 
Fig. 3), the two patches with signs $\pm$ correspond to the regions $z^2+x_4^2\le \ell^2$ 
or $z^2+x_4^2 \ge \ell^2$, respectively, and 
are related to each other by an inversion transformation 
$(z, x_4)\rightarrow \ell^2(z/(z^2+x_4^2), x_4/(z^2+x_4^2))$.  
If we wished and did not insist on the 
conformal gauge, we could have represented the above solution 
using a single patch. In fact, the  
solution with $\alpha=1$ can equivalently be expressed using complex coordinates both for world sheet 
and target space as
\begin{equation}
\tau+i \tilde{\theta}=\ln i{x_4+\ell +i z\over x_4-\ell +iz} , 
\end{equation}
where $\tilde{\theta}$ is related to $\sigma$ by 
$\tanh \sigma=\cos \tilde{\theta}$. The two patches 
correspond to $-\pi/2\leq \tilde \theta \leq 0 $ and 
$0 \leq \tilde{\theta}\leq  \pi/2$. 
The solution with other values of $\alpha$ can 
of course be obtained by conformal 
transformation in the target space. 
This form is very suggestive, since it 
strikingly resembles to similar expressions in the light-cone string 
theory in flat spacetime.  It might be worthwhile to 
elaborate this form further in a direction, say,  
of extending it to cases with multi-point insertions 
of local operators. For the purpose of the present note, however, 
we will use the above two-patch representation.   
\begin{figure}[htbp]
\begin{center}
\begin{minipage}[t]{0.3\textwidth}
\begin{center}
 \includegraphics[width=3cm,clip]{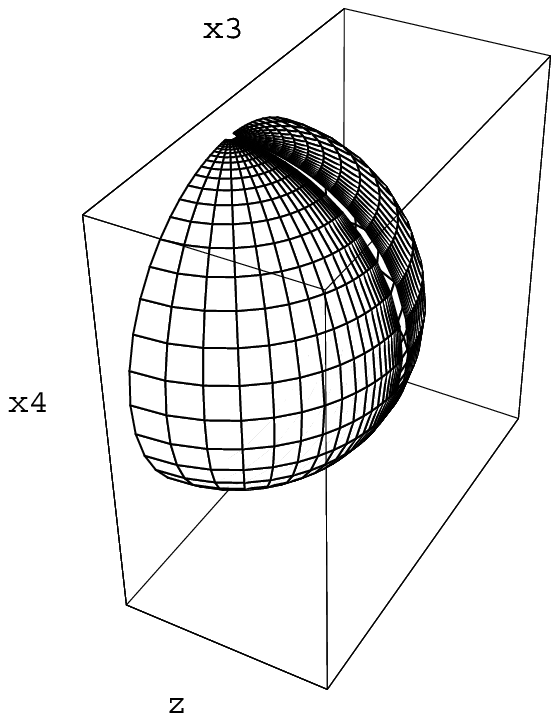}
\caption{{\footnotesize The equal $\tau$-$\sigma$ 
curves on the full world sheet for the case of 
the smallest (with fixed $\ell$)  loop $\alpha=0$.}}
\end{center}
\end{minipage}
\quad
\begin{minipage}[t]{0.3\textwidth}
\begin{center}
\includegraphics[width=3cm,clip]{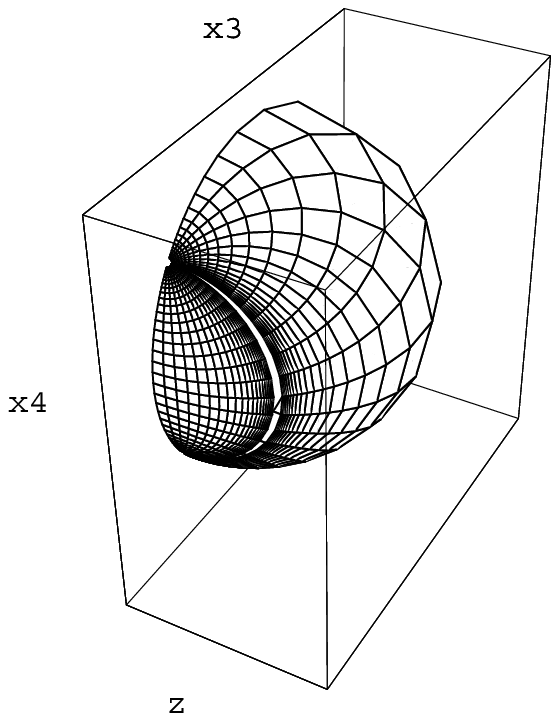}
\caption{{\footnotesize The equal $\tau$-$\sigma$ 
curves on the full world sheet for a generic loop $0<\alpha <1$.}}
\end{center}
\end{minipage}
\quad
\begin{minipage}[t]{0.3\textwidth}
\begin{center}
 \includegraphics[width=3cm,clip]{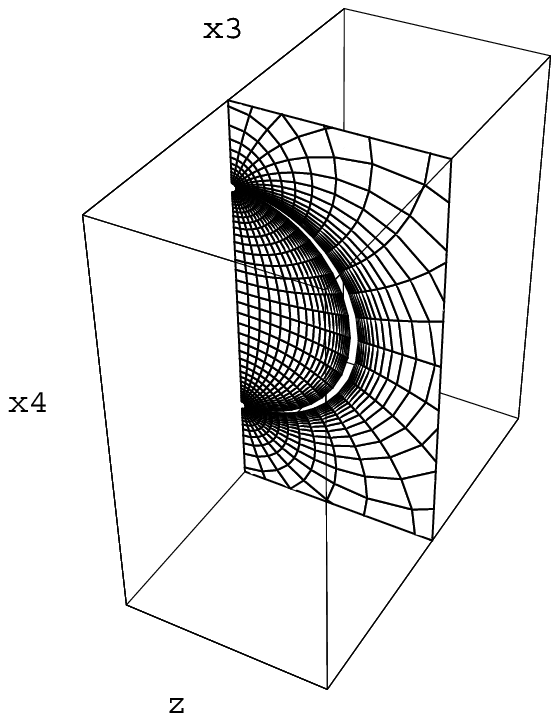}
 \caption{{\footnotesize The equal $\tau$-$\sigma$ 
curves on the full world sheet for the straight-line loop $\alpha=1$.}}
\end{center}
\end{minipage}
\end{center}
\end{figure}

The total R-charge angular momentum carried by this solution 
is 
\begin{equation}
J=\frac{R^2}{\pi \alpha'}\int_0^{\infty} d\sigma 
\tanh^2 \sigma, 
\end{equation} 
which is infinite. Note  that elimination of factor 2 
in the prefactor before the integral  is due to the presence of 
two patches. Thus, we have to actually introduce a cutoff 
for  $\sigma\le \sigma_{\Lambda}$ with a sufficiently large 
$\sigma_{\Lambda}\gg 1$, so that the angular momentum 
is now given as $J\sim R^2(\sigma_{\Lambda}-1)/\pi \alpha'+O(\e^{-2\sigma_{\Lambda}})$. With regard to the dependence 
on $J$, our treatment will be exact up to nonperturbative 
exponential corrections $O(\e^{-2\pi\alpha' J/R^2})$ which can be 
ignored in the power series expansion in $1/J$ 
within our  semi-classical approximation. 

\section{Computing the Wilson-loop expectation value}
Our next task is to evaluate the value of an appropriate classical 
string action functional for this configuration. We have adopted the 
Polyakov-type action for the bulk world sheet and are interpreting the 
process as a scattering event of an open string 
on the  world sheet from 
$\tau=-\infty$ to $\tau=\infty$ for the 
given conserved R-charge angular momentum $J$. 
Thus we should consider the Routhian after a Legendre  
transformation with respect to $\psi$, 
\[
S_{bulk}=\frac{R^2}{4\pi \alpha'}\int d\tau d\sigma
\Big[
\frac{1}{z^2}\Big((\dot{z})^2 +(z')^2 +(\dot{\vec{x}})^2 
+(\vec{x}')^2\Big)
+(\dot{\theta})^2+(\theta')^2
\]
\begin{equation}
\hspace{4cm}
 -\cos^2\theta 
\Big((\dot{\psi})^2 + (\psi')^2  \Big)
- {4 \pi \alpha' \over R^2}\Pi_{\psi}\dot{\psi}\Big]
\end{equation}
with $\Pi_{\psi}=-R^2 \dot{\psi} \cos^2 \theta/2\pi\alpha'$ being the 
$\psi$-momentum. The minus sign of the $\psi$-part comes from 
the Wick rotation. 
Since the world-sheet has a boundary, the action functional
 is subject to ambiguity of boundary terms. We follow the argument of 
ref.\cite{dgo}
which demands that the total action should be a 
functional of momentum  in the radial direction 
of the AdS part, with respect to the 
dependence on the boundary. 
This requirement does not yet 
uniquely fix the choice of the boundary term, 
since they are {\it not} in general invariant under 
canonical transformations. One natural criterion 
for the choice of the variable is that the 
momentum should behave in a well-defined manner as we approach 
the boundary. This criterion is not satisfied if we choose $z$,  
since the radial momentum $\Pi_z=R^2\dot{z}/2\pi \alpha' z^2$ 
for the above solution is not well-defined at $z=0$, 
corresponding to $\sigma=0$. 
Instead of $z$, we choose $u=1/z$ since $\Pi_u=R^2\dot{u}/2\pi 
\alpha' u^2$ behaves well there. 
These two different choices actually
 lead to boundary contributions with 
opposite signs. 
Thus the boundary action we adopt is 
\begin{equation}
S_{boundary}=\frac{R^2}{2\pi \alpha'}\oint_{boundary}\Big(
d\tau \frac{u'}{u} -d\sigma \frac{\dot{u}}{u}
\Big) .
\end{equation}
Then the expectation value of the Wilson 
operator (\ref{dwilson}) is essentially 
$\e^{-S_{bulk}-S_{boundary}}$ up to 
a possible normalization factor, which should not  
depend on the spacetime configuration of the loop $C$ and 
the positions $\vec{x}_{\pm}$ of insertions provided 
 that the string action captures the gauge-theory dynamics of 
 Wilson loops appropriately. 

\subsection{Regularization I}
To carry out a well-defined evaluation of this action integral, 
we have to introduce a definite regularization scheme which 
controls possible infinities arising at least near the boundary 
region $z\sim 0$ 
corresponding to small $\sigma$ and/or large $\tau$. 
The simplest conceivable prescription is to introduce two 
cutoff parameters $\sigma_0$ and $\epsilon$ on the 
world sheet such that 
\begin{equation}
\sigma\ge \sigma_0, \quad \cosh \tau \le \frac{\ell}{\epsilon}
\label{worldsheetcutoff}
\end{equation}
and take the limits $\sigma_0\rightarrow 0, 
\epsilon \rightarrow 0$ afterward.  The second condition 
amounts to setting a lower bound  for the radial coordinate 
$z$ in the target space 
at the points of insertions for sufficiently large $\sigma$, 
$
\lim_{\sigma_{\Lambda}\rightarrow 
\infty} z(\sigma_{\Lambda}, \tau) \ge \epsilon .
$
In the case of local  operators, this is a standard 
short-distance cutoff in applying the GKPW relation. 
We assume that both the bulk and 
boundary actions should be computed for the same finite 
rectangular region defined by these cutoff conditions (\ref{worldsheetcutoff}). 
For the cutoff $\sigma_{\Lambda}$ associated with 
large R-charge angular momentum, it is not necessary 
to introduce any boundary term. 

Using the Virasoro condition 
\begin{equation}
\frac{1}{z^2}(\dot{z}^2 - (z')^2 +\dot{\vec{x}}^2 
-(\vec{x}')^2)=
1 = -\dot{\theta}^2+(\theta')^2 +\cos^2\theta((\dot{\psi})^2
-(\psi')^2)
\end{equation} 
satisfied by the above solution and also  a part of the equations 
of motion, the bulk action is evaluated to be
\begin{equation}
S_{bulk}=\frac{R^2}{\pi \alpha'}\int_{-\tau(\epsilon)}^{\tau(\epsilon)}d\tau
\int_{\sigma_0}^{\sigma_{\Lambda}}d\sigma 
\frac{\dot{z}^2+\dot{\vec{x}}^2}{z^2}
=\frac{2R^2}{\pi \alpha'}\int^{\tau(\epsilon)}_{0}d\tau
(\sigma-\coth \sigma)\Big|^{\sigma_{\Lambda}}_{\sigma_0}, 
\end{equation}
with $\tau(\epsilon)\sim \log (2\ell/\epsilon)$ 
being the upper bound for $|\tau|$ 
determined by  (\ref{worldsheetcutoff}). 
Because of the conformal isometry, the bulk action 
is independent of the radius $r$ of the circle, except for possible 
implicit dependence through the cutoff prescription. 
The boundary action is equal to 
\[
S_{boundary}=\frac{R^2}{\pi \alpha'}\Big[
\int_{0}^{\tau(\epsilon)}d\tau \Big(
-2\coth \sigma_0+\frac{\sinh\sigma_0 \cosh \tau}{
\cosh \sigma_0\cosh \tau +\alpha}
+\frac{\sinh\sigma_0 \cosh \tau}{
\cosh \sigma_0\cosh \tau -\alpha}
\Big)
\]
\begin{equation}
-\int_{\sigma_0}^{\sigma_{\Lambda}}d\sigma 
\Big(
\frac{\cosh \sigma \sinh \tau(\epsilon)}{ 
\cosh \sigma \cosh\tau(\epsilon)+\alpha}
+
\frac{\cosh \sigma \sinh \tau(\epsilon)}{ 
\cosh \sigma \cosh\tau(\epsilon)-\alpha}\Big)
\Big].
\end{equation}
The contribution of the first term $-2\coth \sigma_0$ in the first brace 
just cancels the same but opposite 
contribution from the second term at $\sigma
=\sigma_0\rightarrow 0$ of the bulk action. 
Therefore, the total action is given, in the limit 
where the cutoff parameters must be finally set, by
\[
S_{total}\equiv 
S_{bulk}+S_{boundary}=2J \log\frac{2\ell}{ \epsilon}
-2J -\frac{2R^2 }{ \pi\alpha'}
\]
\begin{equation}
-\frac{R^2 }{ \pi \alpha'}\sigma_0
\frac{2\alpha }{ \sqrt{\cosh^2 \sigma_0-\alpha^2}}
\Big(\arctan \frac{\cosh \sigma_0-\alpha}{ \sqrt{\cosh^2\sigma_0-\alpha^2}} 
-\arctan \frac{\cosh\sigma_0+\alpha}{ \sqrt{\cosh^2\sigma_0-\alpha^2}}
\Big).\label{totalaction}
\end{equation}
Note that we have still kept the small 
cutoff parameter $\sigma_0 \ll 1$ here, 
since the last term has a rather subtle behavior depending 
on the parameter $\alpha$. 
When $\alpha<1$, this reduces in the limit $\sigma_0
\rightarrow 0$ to 
\begin{equation}
S_{total}\Big|_{\alpha <1}=2J \log\frac{2\ell}{ \epsilon}
-2J -\frac{2R^2 }{ \pi\alpha'} . \label{wscutoff_a<1}
\end{equation}
The case $\alpha=1$ of the straight-line Wilson loop,  
however, is special: We obtain
\begin{equation}
S_{total}\Big|_{\alpha=1}=2J \log\frac{2\ell}{ \epsilon}+
\frac{R^2}{ \alpha'}
-2J -\frac{2R^2 }{ \pi\alpha'}. \label{wscutoff_a=1}
\end{equation}
The additional finite piece $R^2/\alpha'$ in this result 
comes from the $(-)$ patch corresponding to the second term 
in the brace of (\ref{totalaction}) which  is singular when $\alpha=1$. 

Let us now discuss the meaning of these results. 
The dependence $d S_{total}/d(2\ell) =
2J/2\ell$ on the distance $2\ell$ of the inserted local 
fields $Z^J$ and $\overline{Z}^J$ shows the 
correct scaling behavior as it should be by the 
conformal covariance property \cite{dk}
of Wilson-loop operators of the type being treated here. 
On the other hand, the dependence on the loop-scale 
parameter $\alpha$ 
takes  the form 
\begin{equation}
S_{total}\Big|_{\alpha=1}-S_{total}\Big|_{\alpha<1}
=\frac{R^2}{ \alpha'}.
\label{alphadiff}
\end{equation}
This behavior is the same as in the case without local-operator 
insertions.  

On the gauge-theory side, as argued in \cite{dg}
for the case of ordinary circular Wilson loops without 
local operator insertions, 
the relation (\ref{alphadiff}) can be interpreted as 
arising from an anomaly associated with the inversion 
conformal transformation between a finite circle and 
a straight line. Their argument seems go through 
to the case with local operator insertions. 
This is consistent with the ladder-graph approximation 
in the limit $R^4/(\alpha')^2 \gg 1$, since 
we can easily check that the contribution of 
ladder-rainbow graphs 
which take into account a planar set of the propagators of scalar 
($\phi_4$) and gauge fields directly connecting points along the loop is 
nonzero only for the case of finite circle $\alpha<1$ and 
is proportional, apart from a power-behaving prefactor, to 
\begin{equation}
I_1(\sqrt{g^2N}\frac{s}{ 2\pi})I_1(\sqrt{g^2N}\frac{2\pi-s}{ 
2\pi})\rightarrow \e^{\sqrt{g^2N}}  , 
\end{equation}
 using the same notation ($R^4/(\alpha')^2=g^2N \gg 1$) as in the 
 reference \cite{sz}. 
The new parameter $s\, \,  (0<s<2\pi)$ denotes the 
coordinate length between 
two local operator insertions measured along the Wilson loop of 
total length $2\pi$. 
Thus the ratio of the case of finite circles to that of straight line 
agrees with (\ref{alphadiff}). The ladder approximation 
is also consistent with the dependence on $\ell$. The validity 
of such an approximation has not been justified from 
first principle, but is not 
unreasonable in view of remaining supersymmetry for 
this type of deformation 
\cite{Zarembo}. For a discussion related to supersymmetry 
of the Wilson loops of our type, we 
refer the reader to \cite{dk, dr}. 

The dependence  on the short-distance cutoff parameter $\epsilon$ 
must be removed by wave-function renormalization of our
 Wilson loop operator  as usual. On the other  hand,  
it is not entirely clear whether we can put any universal 
meaning on the remaining finite contribution 
which does not depend on loop-configuration 
parameters $\ell$ and $\alpha$, but does depend on the 
coupling constant. 
For the purpose of clarifying this problem, 
it is useful to study a different 
regularization scheme. 

\subsection{Regularization II}
\begin{figure}[htbp]
\begin{center}
\begin{minipage}[t]{0.45\textwidth}
\begin{center}
 \includegraphics[width=5cm,clip]{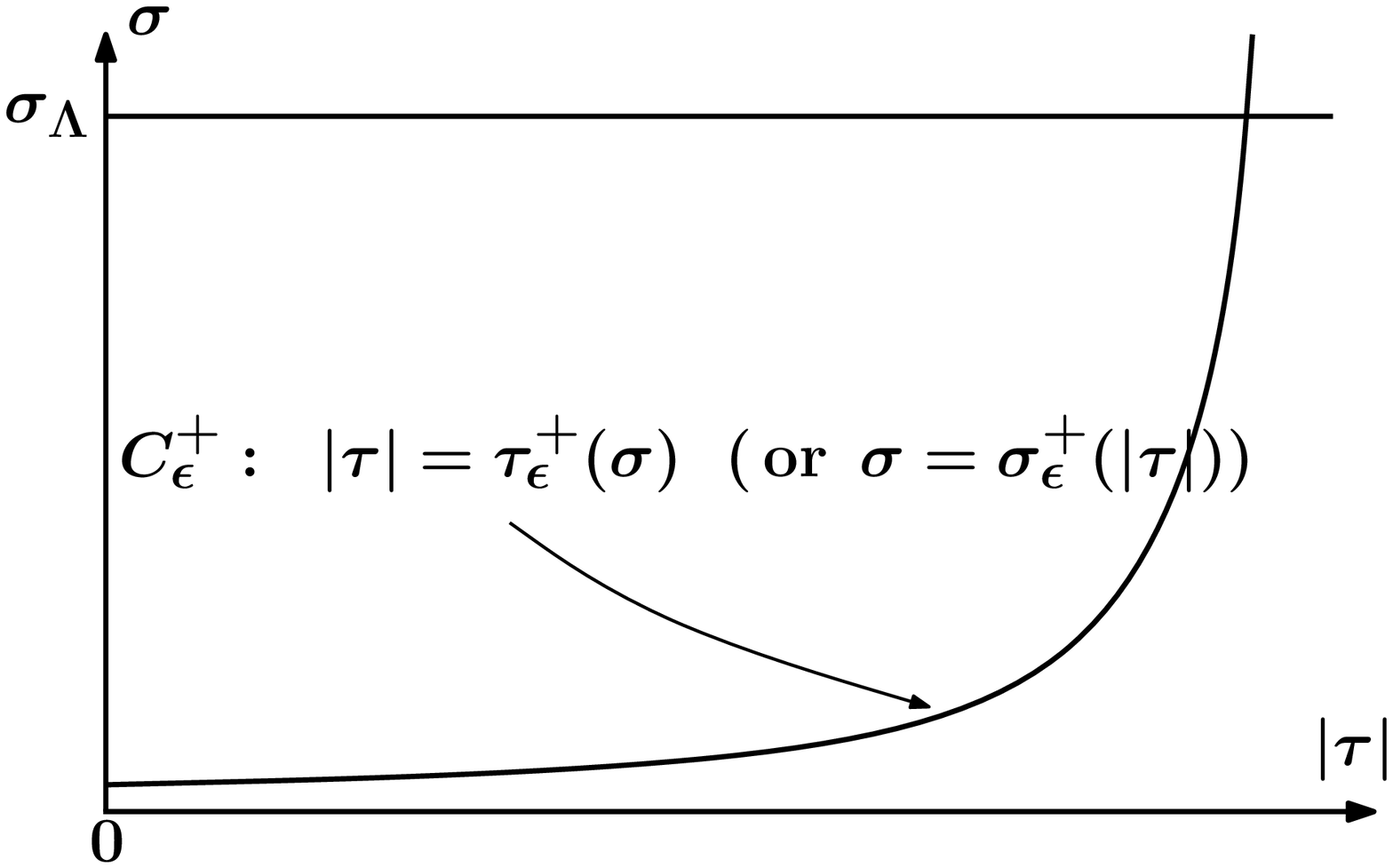}
\caption{\footnotesize{The boundary contours $\sigma=\sigma_\Lambda$ and
 $C_\epsilon^+$ for the ($+$)-patch. 
The functions $\tau_\epsilon^+(\sigma)$ and $\sigma_\epsilon^+(\tau)$
are determined by \eqref{C+} with $\tau > 0$. 
}
}
\label{cutoff+}
\end{center}
\end{minipage}
\qquad
\begin{minipage}[t]{0.45\textwidth}
\begin{center}
 \includegraphics[width=5cm,clip]{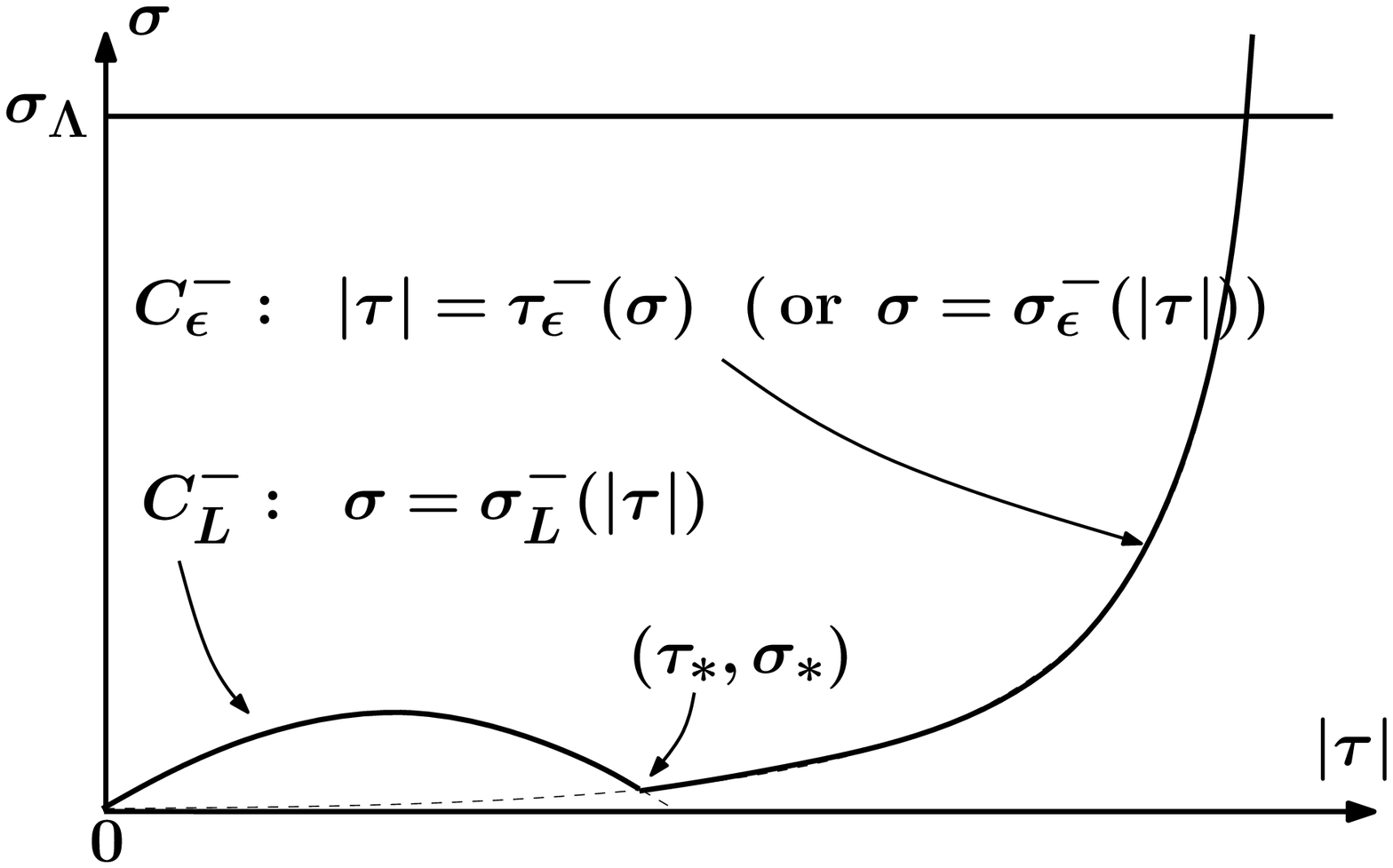}
\caption{\footnotesize{The boundary contours $\sigma=\sigma_\Lambda$,
 $C_\epsilon^-$ and $C_L^-$ for the ($-$)-patch with $\alpha=1$. 
The functions $\tau_\epsilon^-(\sigma)$ and $\sigma_\epsilon^-(\tau)$
are determined by the first equation of \eqref{C-} and 
 $\sigma_L^+(\tau)$ is determined by the second equation
 of \eqref{C-} with $\tau > 0$. }}
\label{cutoff-}
\end{center}
\end{minipage}
\end{center}
\end{figure}

    From the target-space point of view, the infinities of $S_{bulk}$ arise
in two ways: One is from singular behavior of the 
Poincar\'{e} metric
 near the conformal boundary, and  the other is from the 
infinite extension of the world sheet in the direction of $x_4$ 
for $\alpha=1$. Instead of cutoffs in terms of the world-sheet 
coordinates, it is then equally natural to regularize these infinities
using the target-space coordinates as
\begin{equation}
 \big(z(\tau,\sigma)\big)_\pm \geq \epsilon, 
  \quad \big|\big( x_4 (\tau,\sigma) \big)_- \big| \leq L.
  \label{cutoff}
\end{equation}
The second condition is meaningful 
only in the case of $\alpha=1$.
The similar condition for $(x_4)_+$ is always satisfied for $L>\ell$.
The boundaries of the domain of integration with respect to 
the $(\tau, \sigma)$ coordinates 
are given by the following curves 
\begin{align}
 & \textrm{($+$)-patch} \qquad
 C_\epsilon^+
 :\quad \ell \sinh \sigma = \epsilon \big( \cosh \sigma \cosh \tau + \alpha
 \big) \quad(\alpha \leq 1),  \label{C+} \\
 & \textrm{($-$)-patch} \qquad
 \begin{cases}
  C_\epsilon^-: \quad  
  \ell \sinh \sigma = \epsilon \big( \cosh \sigma \cosh \tau - \alpha
 \big) \quad  (\alpha \leq 1),\\
  C_L^-: \quad
  \ell \cosh \sigma |\sinh \tau| = L 
  \big(
  \cosh \sigma \cosh \tau  - 1
  \big) \quad (\alpha = 1).
 \end{cases} 
 \label{C-}
\end{align}
Because of these boundary curves with complicated 
$\tau$-$\sigma$ dependence, 
integrals become more cumbersome than 
the previous regularization. The curves 
with $\alpha=1$ are illustrated
 in Fig. \ref{cutoff+} and \ref{cutoff-}.
In the case of the ($+$)-patch, 
curves with $\alpha < 1$ and $\alpha=1$ are essentially the same. 
As for the case of ($-$)-patch, the curve $C_L^-$ 
in Fig. \ref{cutoff-} does not exist for $\alpha<1$.
Remember that as before we have another cutoff $\sigma_\Lambda$ related to the angular momentum $J$. For definiteness, 
we take the limit $\epsilon \to 0$ and $L \to \infty$ for 
fixed large $J$. It is not 
difficult to check that the opposite order of the limits 
gives the same final result. 

For $\alpha < 1$, contributions from two patches are essentially the 
same. We give brief summary only for the ($+$)-patch. The
bulk contribution is
\begin{equation}
 S_{bulk}^+ 
 = 
 \frac{R^2 }{ 2 \pi \alpha'}
 \int_{-\tau_\epsilon^+(\sigma_\Lambda)}
     ^{\tau_\epsilon^+ (\sigma_\Lambda)} d \tau
   \int_{\sigma_\epsilon^+(\tau)}^{\sigma_\Lambda} d \sigma
  \frac{\dot z^2+ \dot{\vec{x}}^2 }{ z^2}
 =
 \frac{R^2 }{ \pi \alpha'}
 \int_0^{\tau_\epsilon^+(\sigma_\Lambda)} d\tau
 \left(\sigma - \coth \sigma\right)
 \Big|_{\sigma_\epsilon^+(\tau)}^{\sigma_\Lambda} \label{S_bulk^+},
\end{equation}
with $\tau_{\epsilon}^+(\sigma)$ and $\sigma_\epsilon^+(\tau)$ being 
determined by \eqref{C+}.
The boundary contribution is 
\begin{align}
 S_{boundary}^+ = 
 &
 \frac{R^2}{\pi \alpha'}
 \Big[
  \int_0^{\tau_\epsilon^+ (\sigma_\Lambda)} d \tau
 \Big(
  - \coth \sigma_\epsilon^+(\tau) 
  + \frac{
    \sinh \sigma_\epsilon^+(\tau) \cosh \tau 
    }{
    \cosh \sigma_\epsilon^+(\tau) \cosh \tau + \alpha
    } 
 \Big) \notag \\
 &
 -\int_{\sigma_\epsilon^+(0)}^{\sigma_\Lambda} d \sigma
 \Big(
 \frac{
 \cosh \sigma \sinh \tau_\epsilon^+(\sigma) 
 }{ 
 \cosh \sigma \cosh \tau_\epsilon^+(\sigma) + \alpha
 }
 \Big) 
 \Big]. \label{S_boundary^+}
\end{align}
%In \eqref{S_boundary^+},
The first term, $-\coth \sigma_\epsilon^+(\tau)$, 
in the $\tau$-integral cancels the same but opposite term
in \eqref{S_bulk^+} as before, and the remaining 
$\tau$-integral gives $R^2/\pi \alpha'$. 
The $\sigma$-integral in the second line of \eqref{S_boundary^+}
 can be rewritten as 
\begin{equation}
 \frac{R^2}{\pi \alpha'}
  \Big[
- \frac{\epsilon }{ \ell}
  \Big[ 
  \sigma \coth \sigma \sinh \tau_\epsilon^+(\sigma)  
  \Big]_{\sigma_\epsilon^+(0)}^{\sigma_\Lambda}
 + \frac{\epsilon }{ \ell}
  \int_{\sigma_\epsilon^+(0)}^{\sigma_\Lambda} d \sigma \, \sigma
  \frac{d }{ d \sigma} 
  \left( 
    \coth \sigma \sinh \tau_\epsilon^+ (\sigma)
  \right)
  \Big], 
\end{equation}
of which 
the first term gives $-( R^2/\pi \alpha' )\sigma_\Lambda $ and the second term 
goes to zero in the limit $\epsilon \to 0$.
Thus the final result is 
\begin{equation}
 \Big[S_{bulk}^+ + S_{boundary}^+ \Big]_{\alpha<1}
  = 
 J \log \frac{2\ell }{ \epsilon} - J 
 - 
 \frac{R^2 }{ \pi \alpha'} 
 \int_0^{\tau_\epsilon^+ (\sigma_\Lambda)} d \tau \, \sigma_\epsilon^+(\tau).
 \label{S_bulk^++S_boundary^+} 
\end{equation}
Adding the contribution from the other patch, 
we reproduce the same result as in our first regularization scheme,  
except for the last term; we will later estimate the sum of the last term 
in \eqref{S_bulk^++S_boundary^+} and the
similar one from the ($-$)-patch to be a $J$-independent 
 finite constant. 

Next we turn to the case $\alpha = 1$. In contrast to the 
case $\alpha<1$, 
the second cutoff in \eqref{cutoff} plays an important role.
Only the evaluation of the ($-$)-patch needs modification.
The bulk and the boundary contributions are
\begin{align}
 S_{bulk}^-
 & = 
 \frac{R^2}{\pi \alpha'}
 \Big[
 \int_0^{\tau_\ast} d \tau 
 \left(  \sigma - \coth \sigma \right) 
 \Big|_{\sigma_L^-(\tau)}^{\sigma_\Lambda}
 +
 \int_{\tau_\ast}^{\tau_\epsilon^-(\sigma_\Lambda)} d\tau
 \left( \sigma - \coth \sigma \right)
 \Big|_{\sigma_\epsilon^-(\tau)}^{\sigma_\Lambda}
 \Big],
 \label{S_bulk^-} \\[2mm]
 S_{boundary}^-
 & 
 =
 \frac{R^2 }{ \pi \alpha'}
 \Big[
 \int_{C_L^-}
 \Big(
 d \tau
 {u' \over u}
 -
 d \sigma
 {\dot u \over u}
 \Big)
 +
 \int_{C_\epsilon^-}
 \Big(
 d \tau
 {u' \over u}
 -  
 d \sigma
 {\dot u \over u} 
 \Big)
 \Big]. \label{S_boundary^-}
\end{align}
Here, $\tau_\epsilon^-(\sigma)$ and $\sigma_\epsilon^-(\tau)$ are
determined by the first equation in \eqref{C-} 
and $\sigma_L^-(\tau)$ is determined
by the second equation in \eqref{C-}.
The point $(\tau_\ast,\sigma_\ast) $ is where two contours 
$C_\epsilon^-$ and $C_L^-$ with $\tau>0$ meet each other, 
as explained in 
Fig. \ref{cutoff-}, and defines the end-point of the boundary 
integrals in \eqref{S_boundary^-}.   
A little inspection shows that the difference from the case $\alpha<1$
 arises only from the $\tau$-integral along the curve $C_L^-$ in
 \eqref{S_boundary^-}, which is 
rewritten as 
\begin{align}
 \frac{R^2 }{ \pi \alpha'}
 \Big[
 \int_0^{\tau_\ast} d \tau 
 \big(
 - \coth \sigma_L^-(\tau)
 \big)
 +
 \frac{1}{\ell} 
 \int_0^{\sinh \tau_\ast} d x
 \frac{1 }{ x}
 \sqrt{
 -(L^2+\ell^2) x^2 + 2 L \ell \sqrt{1 + x^2} x
 }\;
 \Big].
\end{align}
Here in the second term,
we have used the second equation of \eqref{C-} and changed the variable 
according to $x = \sinh \tau$.
The first term again cancels the corresponding term in
\eqref{S_bulk^-}, and the second term can be evaluated in the limit
$L \to \infty$ and $\epsilon \to 0$ to give $R^2 / \alpha'$.
Summing all the contributions, we obtain
\begin{equation}
 \Big[S_{bulk}^- + S_{boundary}^-\Big]_{\alpha=1}
  =
 J \log \frac{2 \ell }{ \epsilon} - J
  + 
  \frac{R^2 }{ \alpha'}
  -
  \frac{R^2 }{ \pi \alpha'}
  \int_0^{\tau_{\epsilon}^- (\sigma_\Lambda)}
  d \tau \sigma_\epsilon^- (\tau).
  \label{S_bulk^-+S_boundary^-}
\end{equation}
Adding the result for the other patch which is equal to
\eqref{S_bulk^++S_boundary^+},  
we reproduce the same form 
as in the previous prescription, again
up to the last term. In particular, we have reproduced the special 
finite contribution $R^2/\alpha'$ as before. 
In the limit $\epsilon \to 0$,
remaining $\tau$-integrals in \eqref{S_bulk^++S_boundary^+} and 
\eqref{S_bulk^-+S_boundary^-}, whose expressions are valid for any
$\alpha$, can be combined into the following form, apart from the prefactor $R^2/\pi\alpha'$, 
\[
- \lim_{\epsilon\rightarrow 0} \, 
\int_0^{{\rm arc cosh} {{\ell\over \epsilon}} }
  d \tau
  \log
  \left(
   \frac{\ell + \epsilon \cosh \tau }{ \ell - \epsilon \cosh \tau}
  \right),
\]
which dose not depend on $\alpha$.
By expanding the integrand with respect to $\epsilon$ and 
performing the $\tau$-integral order by order,
this is evaluated to be $-\pi^2/4$. 

\section{Conclusion}

To summarize,
 both the behaviors  with respect to $\ell$ and to $\alpha$ do not 
depend on regularizations as expected. However, the 
finite term which is independent of these scale parameters and of $J$ 
actually depends on regularizations.  We can 
conclude that the finite normalization factor, though it depends on the 
coupling constant, cannot be regarded to be universal. 
This is not at all strange, but 
it is desirable to find interpretation for this part from the side of the 
Yang-Mills theory. That would require understanding the 
nature of  nonperturbative 
contributions other than those in simple ladder-type approximations. 
It would also be worthwhile to consider possibility of 
various Ward-like identities 
for the purpose of eliminating ambiguities of regularization. 

Our computations in this note have treated only a particular 
case of Wilson loops with the insertions 
of local scalar operators. Extension to more general spin-chain 
type operators  would be an interesting exercise as in \cite{tsuji}. 
Extension to the case with the insertions of three or more 
local operators must also be a next important problem. Our 
string configurations  
work for sufficiently large R-charge. The case with small but 
nonzero R-charge must perhaps be treated by string configurations 
of different types, in which the density of R-charge angular 
momentum becomes more diffuse as we go inside the string 
world sheet, and hence appropriate solutions for such situation  
should have more complicated time dependence. 
It would also be interesting to study whether we can extend the 
present analysis to cases with insertions 
of operators without R-charge, such as field 
strengths and higher covariant derivatives. 

\vspace{0.2cm}
\noindent
Acknowledgements

We would like to thank Y. Mitsuka, A. Tsuji and especially Y. Kazama
for discussions related to the present subject. 
The work of A. M. is supported in part by JSPS Research Fellowships for Young Scientists.
The work of T. Y. is supported in part by Grant-in-Aid for Scientific Research (No. 13135205 (Priority Areas) and No. 16340067 (B))  from the Ministry of  Education, Science and Culture, and also by Japan-US Bilateral Joint Research Projects  from JSPS.

\end{document}